\documentclass[fleqn,10pt]{wlscirep}
\usepackage[utf8]{inputenc}
\usepackage[T1]{fontenc}
\usepackage{float}

\title{Kidney segmentation in neck-to-knee body MRI of 40,000 UK Biobank participants}

\author[1,*]{Taro Langner}
\author[1]{Andreas Östling}
\author[2]{Lukas Maldonis}
\author[1]{Albin Karlsson}
\author[1]{Daniel Olmo}
\author[2]{Dag Lindgren}
\author[2]{Andreas Wallin}
\author[2]{Lowe Lundin}
\author[1,3]{Robin Strand}
\author[1,2]{H\r{a}kan Ahlstr\"{o}m}
\author[1,2]{Joel Kullberg}

\affil[1]{Department of Surgical Sciences, Uppsala University, 751 85 Uppsala, Sweden}
\affil[2]{Antaros Medical AB, BioVenture Hub, 431 53 Mölndal, Sweden}
\affil[3]{Department of Information Technology, Uppsala University, 751 85 Uppsala, Sweden}

\affil[*]{E-Mail: taro.langner@surgsci.uu.se}

\begin{abstract}
	
The UK Biobank is collecting extensive data on health-related characteristics of over half a million volunteers. The biological samples of blood and urine can provide valuable insight on kidney function, with important links to cardiovascular and metabolic health. Further information on kidney anatomy could be obtained by medical imaging. In contrast to the brain, heart, liver, and pancreas, no dedicated Magnetic Resonance Imaging (MRI) is planned for the kidneys. An image-based assessment is nonetheless feasible in the neck-to-knee body MRI intended for abdominal body composition analysis, which also covers the kidneys.
In this work, a pipeline for automated segmentation of parenchymal kidney volume in UK Biobank neck-to-knee body MRI is proposed. The underlying neural network reaches a relative error of 3.8\%, with Dice score 0.956 in validation on 64 subjects, close to the 2.6\% and Dice score 0.962 for repeated segmentation by one human operator. The released MRI of about 40,000 subjects can be processed within two days, yielding volume measurements of left and right kidney. Algorithmic quality ratings enabled the exclusion of outliers and potential failure cases. The resulting measurements can be studied and shared for large-scale investigation of associations and longitudinal changes in parenchymal kidney volume.

\end{abstract}

\begin{document}

\flushbottom
\maketitle
%
%
\thispagestyle{empty}


\vspace{-0.5cm}
\section{Introduction}

The UK Biobank studies more than half a million volunteers participants, collecting health data on medical records, blood and urine samples, lifestyle, and genetics.\cite{sudlow_uk_2015} Together with the vast range of metadata and a long-term follow-up, medical images are acquired for a subgroup of 100,000 participants. Of these, 10,000 are also scheduled to attend a repeat scan at a second, later imaging visit. The protocols include Magnetic Resonance Imaging (MRI) of the brain, heart, pancreas and liver, but also neck-to-knee body MRI\cite{west_feasibility_2016} which combines vast amounts of anatomical information in a single comprehensive 6-minute scan, which covers the all tissue of the kidneys in overlapping imaging stations.
The human kidney plays a vital role in the filtration of blood, secretion of hormones, and regulation of blood pressure. Its shape and function are impacted by genetic factors, but also underlie natural variation based on sex, body size, and age.\cite{emamian1993kidney} In addition to congenital anomalies such as renal fusion, or horseshoe kidneys,\cite{glodny2009kidney} and autosomal dominant polycystic kidney disease (ADPKD),\cite{sharma2017kidney} morphological changes with associated medical complications can arise from factors such as chronic kidney disease, hypertension, \cite{hoy2008nephron} and diabetes. \cite{rossing2002risk} Kidney volume as a biomarker is therefore of clinical interest for diagnostics, monitoring of disease progression, and medical hypothesis testing. With the extensive image data available in the UK Biobank, non-invasive, image-based assessments of kidney volume could provide a substantial sample size of these measurements.

In clinical practice, kidney volume is often approximated with a rotational ellipsoid model based on kidney width, depth, and length as obtained by sonography.\cite{bakker1998vitro} However, validation with water displacement methods has shown that ellipsoid models can underestimate kidney volume by up to 29\%,\cite{cheong2007normal} or 25\% even with MRI.\cite{bakker1998vitro} As an alternative, measurements by voxel count, or disc-summation, can be obtained by delineation of three-dimensional voxels which correspond to kidney tissue in volumetric medical images. When obtained from MRI, these segmentation-based measurements have been found to show no significant deviation to those determined by water displacement.\cite{bakker1998vitro} When applied to the UK Biobank, manual segmentation is no longer feasible, however, as even a typical processing time of ten minutes per subject would amount to tens of thousands of man-hours for the UK Biobank cohort as a whole. 
A rich body of literature has been devoted to computer-aided segmentation techniques of the kidneys and other visceral organs in volumetric medical imaging data. For the kidney in particular, various approaches have been proposed predominantly for image data from Computed Tomography (CT), including techniques based on statistical shape models and region growing \cite{lin2006computer}, graph cuts \cite{ali2007graph}, and deformable boundaries \cite{shehata20183d}. 
Contemporary benchmark challenges are increasingly dominated by machine learning techniques such as deep learning with convolutional neural networks, as seen in the MICCAI \textit{2019 Kidney and Kidney Tumor Segmentation (KiTS19)} \cite{heller2019state} and with CT image data, in which similar approaches have also been proposed for measurements of total volume in subjects with ADPKD. \cite{sharma2017automatic} Fully convolutional networks for semantic image segmentation \cite{long2015fully} range from architectures such as the U-Net with 2D data\cite{ronneberger2015u} to 2.5D\cite{han2017automatic} and 3D techniques,\cite{isensee2019attempt} which are able to learn the task of segmenting specific image structures from reference data in training.
In the UK Biobank, related approaches have already been applied for segmentation of cardiac MRI of up to 5,000 subjects \cite{zheng20183, bai2018automated} and large-scale segmentation of this data has been conducted with other methods as well, such as sparse active shape models on up to 20,000 subjects. \cite{attar2019quantitative}

The purpose of this work is to propose, validate, and apply a segmentation pipeline for automated quantification of parenchymal kidney tissue in UK Biobank neck-to-knee body MRI. A neural network based on a 2.5D U-Net variation is evaluated in cross-validation and applied for inference to all 40,000 subjects with available MRI data, resulting in measurements of healthy tissue volume of the left and right kidney, as well as their mutual distance. Potential failure cases and other outliers are identified with algorithmic quality ratings, with a large number of anomalies such as renal fusion and polycystic cases being highlighted for scrutiny. We are not aware of any existing kidney volume measurements within the UK Biobank, or any other study with MRI-based measurements of kidney volume at a comparable sample scale. The obtained values and code samples can be shared as return data and made available for further research.

\section{Methods}

A neural network was trained for semantic segmentation of two-dimensional, axial slices of the UK Biobank neck-to-knee body MRI. Manually created reference segmentations of parenchymal kidney volume in 122 subjects were used to train and validate the network, and to make design choices regarding data pre-processing and hyperparameter selection. The resulting network configuration was then embedded in a processing pipeline and applied in inference to the entire cohort, with algorithmic quality ratings flagging suspected failure cases for exclusion. A schematic overview over the experiments is shown in Fig.\ref{fig_overview}.

\begin{table*}[b!]
	\begin{minipage}[t]{0.55\textwidth}
		\begin{figure}[H]
			\centering
			\includegraphics[width=\textwidth]{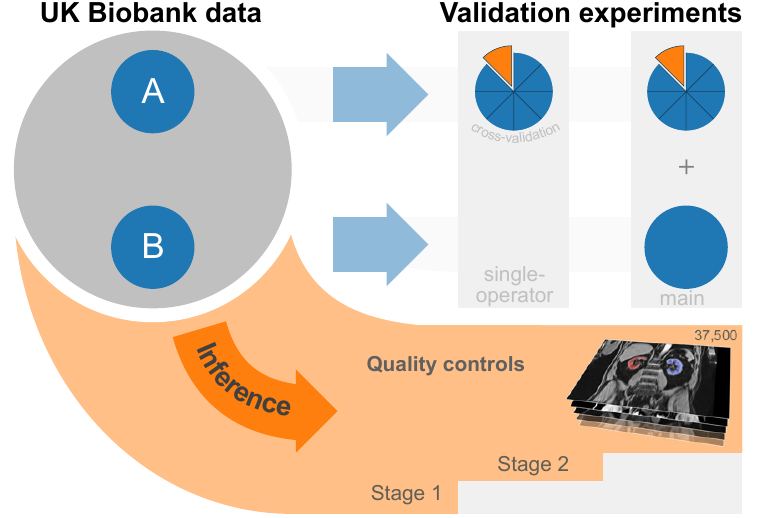}	
			\caption{Among all UK Biobank subjects, two subsets \textit{A} and \textit{B} were manually segmented. A neural network was evaluated in two cross-validation experiments on these and applied for inference to all remaining subjects. After excluding 5\% of results in two quality control stages, about 37,500 measurements remain as final result.} 
			\label{fig_overview}
			
		\end{figure}
	\end{minipage}
	\hfill
	\begin{minipage}[t]{0.45\textwidth}
		\begin{figure}[H]
			
			\includegraphics[width=\textwidth]{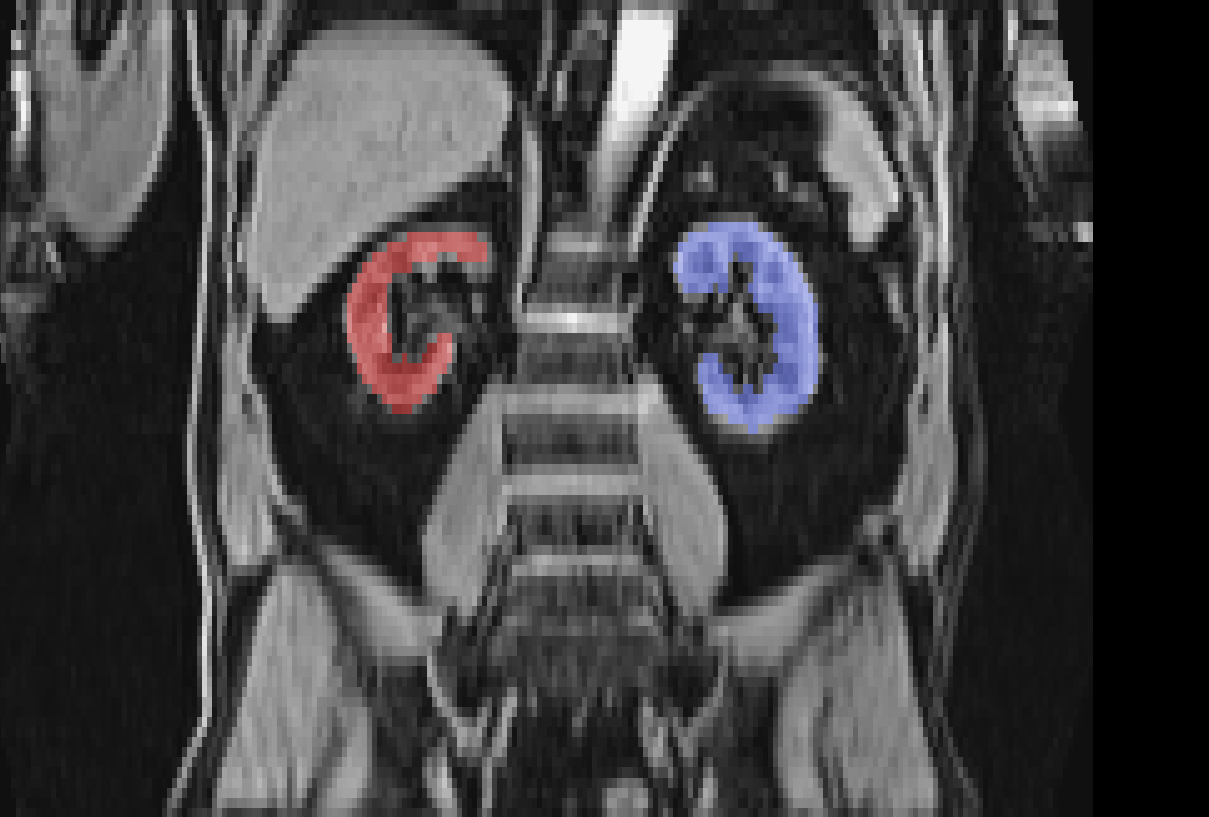}	
			
			\caption{Segmented parenchymal tissue of right (red) and left (blue) kidney in MRI of a male subject.}
			
			\label{fig_segmented}	
		\end{figure}
	\end{minipage}
\end{table*}
\subsection{UK Biobank data}

At the time of writing, UK Biobank neck-to-knee body MRI of 40,264 participants has been released. Subjects were recruited by letter from the National Health Service and scanned at three different imaging centres in Great Britain with a Siemens Aera 1.5T device, using a dual-echo protocol that acquired overlapping images in six stations covering the body from neck to knee within about 6 minutes with TR = 6.69, TE = 2.39/4.77 ms, and flip angle 10deg.\cite{west_feasibility_2016} The reconstructed, volumetric station images encode voxel-wise intensity values with a separate water and fat signal (UK Biobank field 20201-2.0). The head, arms, and lower legs extend outside of the field of view and are often distorted near the image borders. 

The kidneys are typically located in the second and third imaging station, each of which were acquired in a 17s breath-hold with typical dimensions of \mbox{(224 $\times$ 174 $\times$ 44)} voxels of \mbox{(2.232 $\times$ 2.232 $\times$ 4.5)} mm.
In this work, those subjects with image artefacts in these two stations, such as water-fat swaps, background noise, metal objects, but also non-standard poses, misalignment in the scanner, and corrupted data were excluded after visual inspection of mean intensity projections,\cite{Langner2020} leaving 39,560 subjects. At scan time, these men and women (52\% female) were 44 to 82 (mean 64) years old, with BMI 14 to 62 (mean 27) $kg/m^2$ and a 95\% majority of self-reported White British ethnicity.

\subsubsection{Reference segmentations}

Three operators created reference segmentations by marking all voxels corresponding to healthy, parenchymal kidney tissue in the water signal of the second and third imaging station. The segmented tissue corresponds to the cortex and medulla, both of which appear with high MRI water signal intensity. Based on their lower signal intensities cysts, the calyces, ureters, and major vessels were excluded. These reference segmentations were used to train and validate the neural network. For the final measurements, the left and right kidney were separated by subsequent post-processing with an algorithmic connected component analysis. An example for a segmented MRI slice is seen in Fig.\ref{fig_segmented}.

Dataset \textit{A} consists of 64 subjects selected by random sampling stratified by age, gender, and weight,\cite{ostling2019automated} whose water signal images were manually segmented in the software SmartPaint\cite{malmberg2017smartpaint} by an experienced image analyst. The consistency of these segmentations was evaluated on a subset of 5 subjects, which were segmented repeatedly by the operator for a blinded assessment of intra-operator variability.

Dataset \textit{B} contains another 64 subjects, with no overlap to dataset \textit{A}, with 33 cases segmented by the second and 31 cases by the third operator, also segmented in SmartPaint. Instead of using a fully manual procedure, this dataset was segmented by correcting the proposals generated by a preliminary inference network trained on dataset \textit{A}. The 64 candidates were selected among the most challenging cases based on an algorithmic quality rating for segmentation smoothness which is described in more detail further below. As a result, subjects with morphological anomalies are over-represented in this dataset, with several pathological cases that are challenging to segment even for human operators. To determine inter-operator variability, these two operators also segmented the same subset of 5 subjects from dataset \textit{A} for which the intra-operator variability was previously determined.

\subsection{Neural network configuration}

A fully convolutional neural network was trained for semantic image segmentation of axial slices. The underlying architecture is a 2.5D variation of the U-Net \cite{ronneberger2015u} with a VGG11 encoder pretrained on ImageNet,\cite{ternaus} extended with ResNet-style short skip connections. The 2.5D input formatting combines three adjacent slices to form one sample, providing additional volumetric information to the network. Similar techniques with five slices \cite{han2017automatic} ranked among the most successful contributions for segmentation of liver tissue in the 2017 MICCAI Liver Tumor Segmentation (LiTS) Benchmark Challenge. \cite{bilic2019liver}

For network training, the three outermost axial slices of each station were removed due to excessive artefacts, such as signal loss and folding. Each remaining axial slice was individually normalized after clipping of the brightest one percent of values for stability.
No image fusion was performed at this stage. The network assigned pixel-wise labels to two-dimensional, axial slices based on a 2.5D input sample is formed by a stack of three adjacent slices from the water signal volume. In addition to the target slice, one additional slice is extracted from above and below each, using a periodic border condition. For an evenly divisible size, the slices were symmetrically zero-padded to form a stack of $224 \times 192 \times 3$ pixels. Each of these stacks forms one input sample for the network, which predicts a two-dimensional segmentation for the central slice in the format of $224 \times 192$ pixels. The network architecture was trained for 80,000 iterations with a pixel-wise cross-entropy loss, batch size one, and online augmentation with randomized, elastic deformations.\cite{ostling2019automated} Using the Adam optimizer, a learning rate of $0.0001$ is maintained until iteration 60,000 and then lowered by factor 10 for improved stability. After reverting the slice padding, the segmented slices with pixel-wise labels can be stacked to obtain voxel-wise labels for an entire input station. A GitHub implementation is linked in the supplementary material.

\subsubsection{Training data}

Three experiments were conducted with this neural network configuration. For training and validation of the network, both dataset \textit{A} and \textit{B} were available, with image data of 64 subjects each. The samples of dataset \textit{B}, however, were selected among the most challenging, including some pathological cases and are largely based on refined segmentations originally proposed by the network itself. Using these samples for validation would yield results that are not representative for the UK Biobank cohort as a whole and dataset \textit{B} was therefore never used for validation. Six of its 64 cases were furthermore excluded due to excessive morphological anomalies, tumours, cysts and congenital renal fusion where both kidneys are interconnected and form a single structure. Thus, 58 cases of dataset \textit{B} remained for further use. 

A single-operator validation was performed by conducting a classical 8-fold cross-validation on the 64 cases of dataset \textit{A}. This dataset was consequently split on subject level into 8 subsets of even size, for each of which in turn segmentations were predicted by a network instance trained on data of all remaining 7 subsets. Each network instance was thereby trained on data of 56 subjects, corresponding to about 4,250 labelled slices. This single-operator cross-validation aims to quantify how well the operator of dataset \textit{A} can be emulated on a representative sample of the UK Biobank.

Secondly, the main validation experiment quantifies the benefit of access to both datasets (\textit{A} $\cup$ \textit{B}) by repeating the cross-validation with the exact same splits, but with samples of dataset \textit{B} added for training only. In this way, the network instance for each split used the same validation subjects as before, but was trained on both the remaining 56 cases of dataset \textit{A} and all well-formatted 58 cases of dataset \textit{B} combined, for a total of 114 subjects, or 8,650 labelled slices for training. The network thereby learns a compromise in segmentation style between all operators, with validation results that are representative for the actual inference pipeline.

Finally, the network was applied for inference itself to all those subjects with no reference data. It was trained on a combined dataset of the 64 cases of datasets \textit{A} and the well-formatted 58 cases of dataset \textit{B}, for a total of 122 cases with about 9,250 labelled slices in total.

\subsection{Measurements}

The second and third stations of a given subject were labelled by the neural network and subsequently fused into a single, combined volume for both the water signal and voxel-wise labels each, by resampling to a common voxel grid and interpolation along the overlapping areas. A kidney volume measurement was obtained from these fused segmentation images by summing up the number of voxels labelled as kidney tissue, scaled with the physical voxel dimensions. Post-processing extracted the two largest connected components individually, which are assumed to be the left and right kidney, identified by the relative position of their centres of mass. The latter also enables a measurement of the relative position and euclidean distance between both kidneys.

\subsection{Validation metrics}

When validating the network output against known reference segmentations, the segmentation quality was evaluated with the Sørensen–Dice coefficient, or Dice score, and Jaccard index. To avoid averaging with empty imaging stations, these metrics were only calculated after fusing the image stations for a given subject.
All measurements were likewise only derived after image fusion, and evaluated with several complementary metrics. Averaging the absolute differences between predicted value and reference for all subjects yields a mean absolute error (MAE). In addition to this value, a relative error measurement is reported for a better sense of scale. Dividing the absolute differences on a per-case basis by the true measurement value, estimated here as the mean between prediction and reference, before averaging, results in a symmetric mean absolute percentage error (SMAPE). For a single example case with true volume of 250cm$^3$, an absolute difference of 25cm$^3$ would thereby amount to a SMAPE of 10\%. Instead of estimating the true value as the mean of prediction and reference, an alternative would be to simply use the reference value directly. However, the known high variation between references created by different operators suggests that the chosen symmetrical definition may be more robust. In addition to these metrics, the quality of fit between predicted values and reference can be quantified with the coefficient of determination (R$^2$), whereas error bounds are estimated by the 95\% limits of agreement (LoA).

\subsection{Algorithmic quality controls}

When applying the network in inference to those cases with no existing reference measurements, the aforementioned validation metrics can not be calculated. Exhaustive quality control by manual inspection is likewise hardly feasible at this scale. The evaluation during inference is therefore based on algorithmic quality ratings as simple indicators for outliers and potential failure cases. While ratings of high quality provide no guarantee for correct results, low ratings can help to identify those cases that are likely to contain anomalies or potential segmentation failures. The distribution of ratings were examined in two separate control stages, after each of which the most severe outliers were flagged for exclusion (see supplementary material for details). 

The first stage of quality controls evaluates the image quality with an \textit{image fusion rating}, \textit{segmentation fusion rating}, and \textit{location rating}. Even a hypothetically perfect segmentation can result in faulty measurements if parts of the kidney are not contained in the image at all or occur redundantly due to motion. The agreement between both stations in their overlapping area is therefore examined, both for the MRI water signal and their segmented labels.
Large differences indicate bad anatomical alignment between the imaging stations, leading to low quality ratings. Additionally, the relative offset along the longitudinal axis between the centre line of the fused subject volume and the centre of mass of all segmented voxels is penalized. Low values for this rating indicate that the kidneys are located at the top or bottom edge, possibly extending beyond the field of view.

The second quality control stage rates the segmentation quality with a \textit{segmentation smoothness rating} and \textit{scrap volume rating}, examining the smoothness of the segmented volume along the longitudinal axis and the share of voxels which are not part of either of the two largest connected components. The slice-wise segmentation by the 2.5D neural network may encounter failure cases where entirely disconnected islands of tissue are spuriously segmented or excluded due to their position and local appearance. Low ratings indicate atypical shapes for the segmented kidney tissue which may require further scrutiny.

\newpage
\section{Results}

\subsection{Network validation}

\begin{figure}[t]
	\centering
	\includegraphics[width=\textwidth]{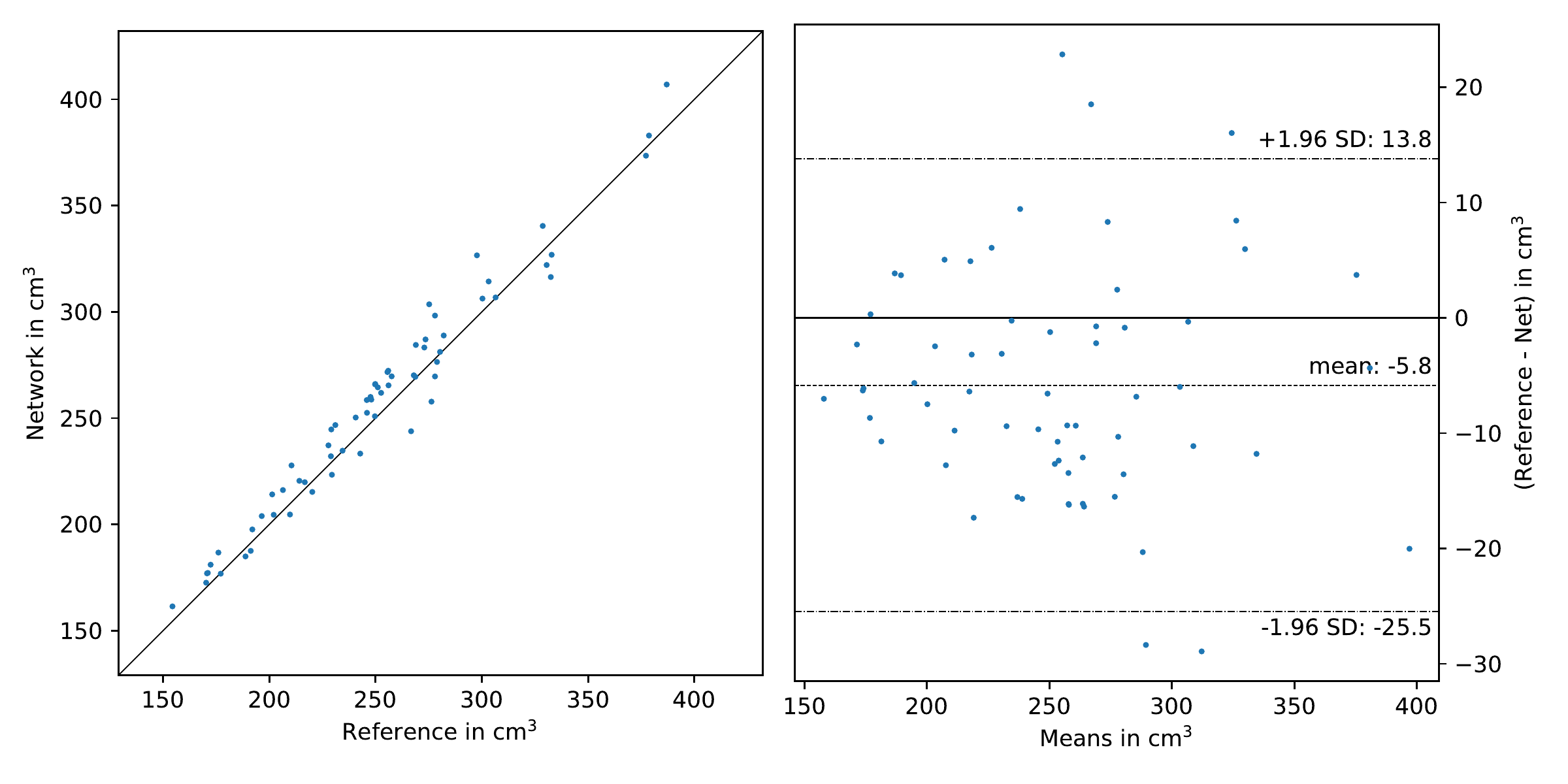}	
	\caption{Main validation result for 64 subjects of dataset \textit{A}, with images from dataset \textit{B} added for training. The diagonal line in the scatter plot on the left represents a hypothetical perfect result, whereas the dashed lines in the Bland-Altman plot on the right give the 95\% Limits of Agreement. When compared to the reference, the network appears to emulate a tendency towards oversegmentation which is also seen in Table \ref{tab_validation} for the operators who provided reference segmentations for \textit{B}.} 
	\label{fig_cv_64}
	\vspace{-0.5cm}
\end{figure}

In both validation experiments, the neural network reached a Dice score of 0.956 on the 64 subjects dataset \textit{A}. The main result, in which the 58 selected subjects of dataset \textit{B} were added for training, measured combined kidney volume with an average error of 10 cm$^3$, or 3.8\%. These values are slightly worse than those achieved by the single-operator cross-validation, and the LoA indicate systematic oversegmentation by the network relative to the operator of dataset \textit{A}, similar to the operators who supplied the training data for dataset \textit{B}. Table \ref{tab_validation} and Fig. \ref{fig_cv_64} show more validation metrics for these results, together with the variability between human operators for context. Additional detail is given in the Supplementary Tables \ref{tab_supp_jaccard}, \ref{tab_supp_cv} and \ref{tab_supp_operators}, with the corresponding Jaccard indices and individual measurements of left and right kidney for both network and operators.

\begin{table}[h]
	\begin{center}
		\caption{Validation results and human variability for combined kidney volume}
		\label{tab_validation}		
		\begin{tabular}{l|rc|rrcc}
			& N & Dice & MAE & SMAPE & R$^2$ & LoA \\
			\hline	
			\textbf{Network validation} &&&&&& \\
			Main result & 64 & 0.956 & 10 cm$^3$ & 3.8\% & 0.950 & (-26 to 14 cm$^3$) \\
			Single-operator & 64 & 0.956 & 9 cm$^3$ & 3.4\% & 0.950 & (-22 to 23 cm$^3$) \\
			
			&&&&&&\\
			\textbf{Human variability} &&&&&&\\
			Intra-operator & 5 & 0.962 & 6 cm$^3$ & 2.6\% & 0.994 & (-4 to 13 cm$^3$) \\			
			Inter-operator & 5 & 0.920 & 27 cm$^3$ & 10.0\% & 0.839 & (-59 to 5 cm$^3$)\\	
			\hline
			\hline
			\multicolumn{7}{l}{Validation on N subjects for the neural network and repeat segmentation by human operators.}\\ 
			\multicolumn{7}{l}{ Whereas the single-operator validation is a classical cross-validation on dataset \textit{A}, the main result} \\
			\multicolumn{7}{l}{ was obtained by training on samples of both datasets \textit{A} and \textit{B}. The resulting measurements of combined}\\
			\multicolumn{7}{l}{ kidney volume were evaluated with the mean absolute error (MAE), symmetric mean absolute} \\ 
			\multicolumn{7}{l}{ percentage error (SMAPE), coefficient of determination (R$^2$) and 95\% limits of agreement (LoA).} \\
			\multicolumn{7}{l}{The latter are calculated as (reference - predicted), yielding a negative shift for oversegmentation.}
		\end{tabular}
	\end{center}
\end{table}



\subsection{Inference}

\begin{figure}[H]
	\centering
	\includegraphics[width=0.33\textwidth]{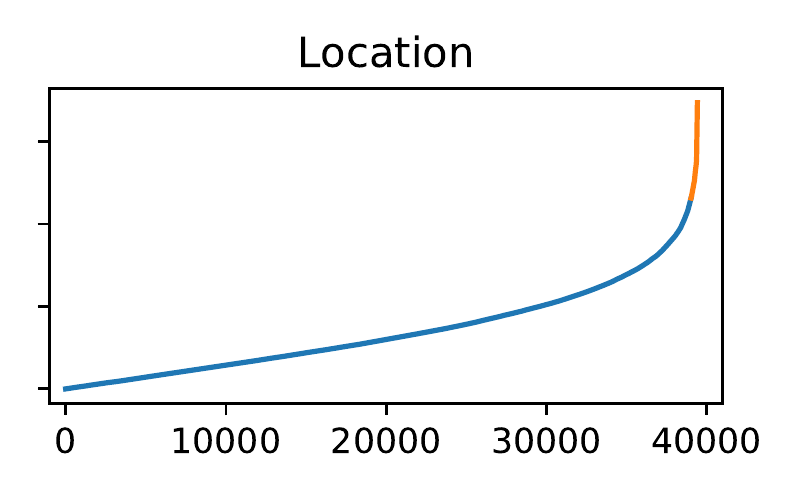}	
	\includegraphics[width=0.33\textwidth]{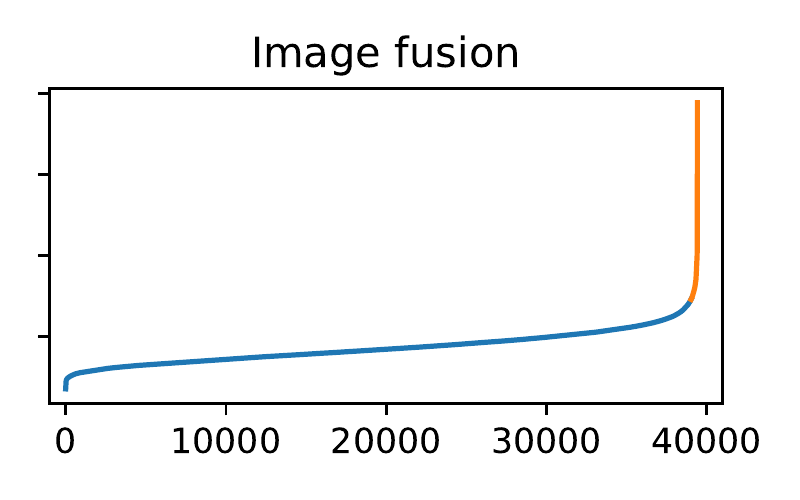}	
	\includegraphics[width=0.33\textwidth]{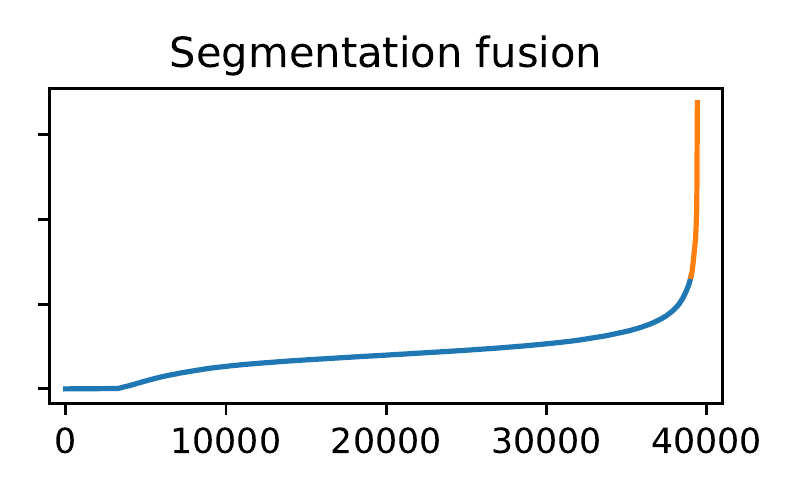}	
	
	\includegraphics[width=0.33\textwidth]{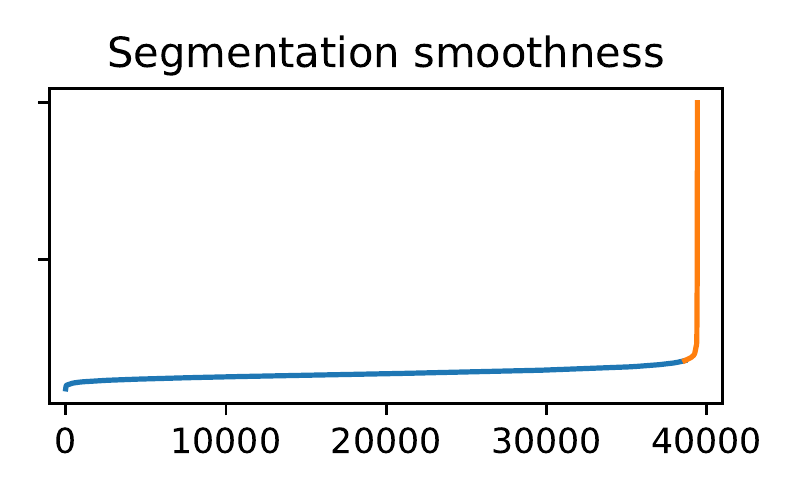}	
	\includegraphics[width=0.33\textwidth]{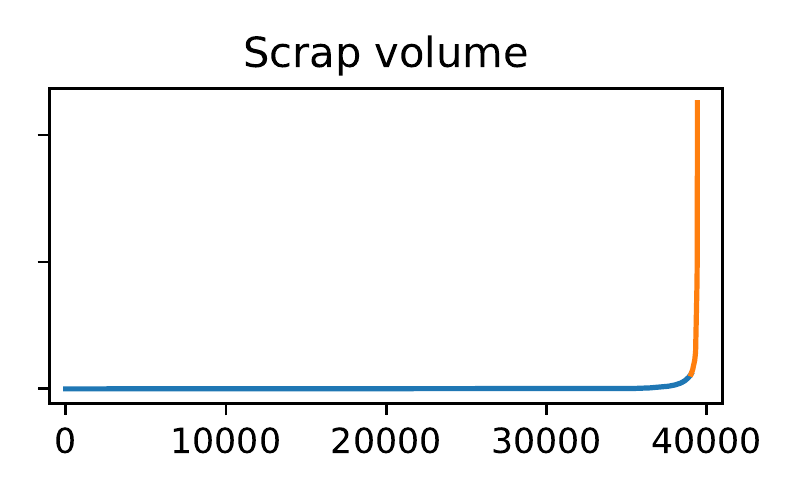}	
	\caption{Distribution of algorithmic quality ratings, sorted separately for each rating. High values of the cost terms indicate low image quality. In stage one (top row) and stage two (bottom row) of quality controls, the highlighted top one or two percent of subjects were accordingly flagged for exclusion as potential failure cases.}
	\label{fig_exclusion_dist}	
\end{figure}

The inference network generated measurements for all those 39,432 subjects lacking reference segmentations. Only a small number of these cases exhibited disjunct or fragmented segmentations, with the \textit{scrap volume} rating indicating that, on average, only about 1 in 900 voxels were not part of the two largest connected components segmented for the given subject (about half of a preliminary run trained on dataset \textit{A} only). Low quality ratings are concentrated in a small subgroup of subjects, as shown in Figure \ref{fig_exclusion_dist}, which were isolated in the following quality control stages.

Based on the algorithmic ratings for image quality, the top one percent of worst \textit{location cost} and \textit{image fusion cost} as well as the top two percent of worst \textit{segmentation fusion cost} were flagged for exclusion in a first control stage. Due to their mutual overlap, the subjects marked in this way amount to about 3.6\% of all cases, many of which show signs of severe motion artefacts or misalignments of the field of view. Some of these cases were trivially re-included, having been flagged by the \textit{location cost} for proximity to the image borders while being too small to extend beyond them.
Next, the algorithmic ratings for segmentation quality were examined for the remaining subjects As the top one percent of worst \textit{segmentation smoothness cost} and worst \textit{scrap volume cost}, another 1.8\% of subjects were flagged in this step. More cases with motion were caught at this stage, as well as genuine failure cases in which the network mistakenly segmented parts of the spleen or liver, but also cases of fragmentation caused by severe cystic formations. In total, 5\% of subjects were ultimately excluded, with representative cases shown in Supplementary Fig. \ref{fig_supp_excluded}. Many of these cases contain pathological anomalies. In turn, perhaps up to a third of them could potentially be re-included without any corrections, but were not considered any further in this work.

\begin{figure}[t]
	\centering
	\includegraphics[width=\textwidth]{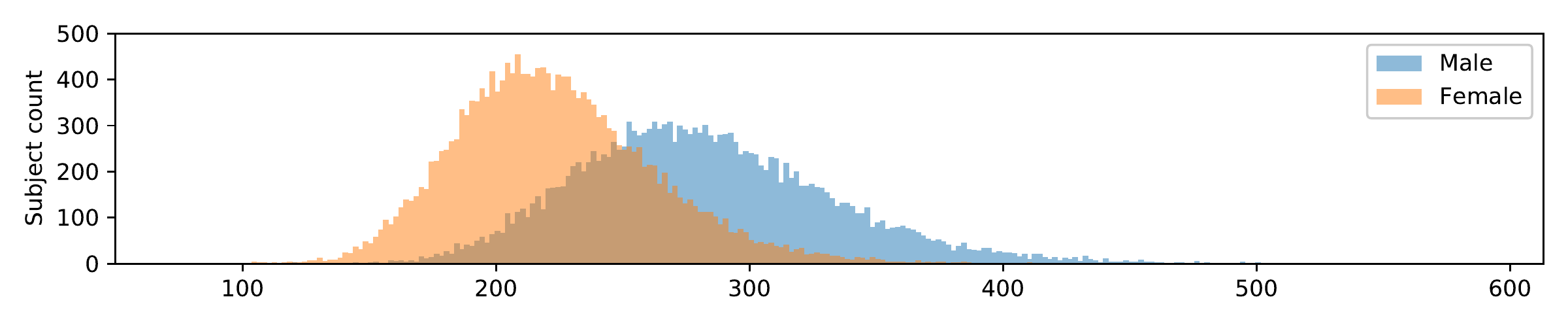}	
	\caption{Inferred UK Biobank parenchymal kidney volume (left + right) in cm$^3$ for 17,846 male and 19,622 female subjects.}
	\label{fig_inference_hist}	
\end{figure}

\begin{table*}[b!]
	\begin{center}
		\caption{Inferred UK Biobank parenchymal kidney volumes in cm$^3$}
		\label{tab_inference_results}		
		\begin{tabular}{l|cl@{\hskip 0.1cm}c|cr@{\hskip 0.2cm}l}
			
			Property & mean $\pm$ SD & [min, & max] & median & (10\%, & 90\%) \\
			\hline	
			\textbf{Combined} &&&&&&\\
			male  & 282 $\pm$ 51 & [91, & 586]& 277 & (221, & 348) \\
			female & 224 $\pm$ 40& [76, & 499] & 220 & (177, & 276) \\	
			
			&&&&&&\\		
			\textbf{Left} &&&&&&\\
			male & 143 $\pm$ 29& [0, & 408]  & 141 & (110, & 178) \\
			female & 114 $\pm$ 22 & [0, & 304] & 112 & (88, & 141) \\
			
			&&&&&&\\	
			\textbf{Right} &&&&&&\\
			male & 139 $\pm$  28& [0, & 408]  & 137 & (108, & 173) \\
			female & 110 $\pm$ 22 & [0, & 268] & 108 & (86, & 138) \\

			\hline
			\hline
			\multicolumn{7}{l}{Male (N=17,846) and female (N=19,622) parenchymal kidney volumes} \\
			\multicolumn{7}{l}{ in cm$^3$. SD denotes the standard deviation.} \\
		\end{tabular}
	\end{center}
\end{table*}
After these exclusions, 37,468 subjects remain, with 17,846 men and 19,622 women. Disjunct scrap volume occurs in about 20\% of these subjects, but amounts to only 1 in 2,200 segmented voxels on average and never exceeds a share of 2.5\% for any individual. Outliers with unusually high or low volumes were inspected as potential failure cases, but were found to be plausible measurements of subjects with missing kidneys, unilateral hypertrophy/atrophy, or were associated with outliers of body size. An in-depth medical analysis of the resulting measurements is beyond the scope of this work and remains to be explored in the future. However, as a brief summary, Fig. \ref{fig_inference_hist} shows the distribution of measured combined kidney volume, with additional statistics given in Table \ref{tab_inference_results}, and further detail on the offset between kidneys in Supplementary Table \ref{tab_inference_offset}.

\subsection{Runtimes and memory requirements}

Training the 2.5D U-Net on an Nvidia RTX 2080 Ti 11GB GPU for 80,000 iterations required about 30 minutes per split, or about 3.5 hours for the entire 8-fold cross-validation. The MRI data for water and fat signal was stored in DICOM format on an encrypted USB-SSD, amounting to 750 GB for 40,000 subjects. The inference pipeline loaded and processed individual scan volumes from this drive. Despite an efficient GPU implementation, the image fusion formed the bulk of processing time during inference, amounting to almost 30 hours for all 40,000 subjects.

\section{Discussion}

As main validation result, the proposed measurements for total kidney volume agree with the reference for a mean error of \mbox{10 cm$^3$}, or 3.8\%, and Dice score 0.956. The assessment of human performance showed slightly superior results for  blinded repeat segmentation by a single operator, with mean error \mbox{6 cm$^3$}, or 2.6\%, and Dice score 0.962, whereas the variability between different human operators was more than twice as large as the network error. When applied to the entire cohort, around 37,500 subjects yielded volume measurements with no signs of potential measurement failure, whereas another 5\% require further controls. 

Only healthy parenchymal tissue was segmented, including cortex and medulla while excluding the renal pelvis, calyces, ureters, major vessels and cysts. The measurements obtained by the inference therefore differ from those typically used for the tracking of conditions such as ADPKD, which may nonetheless benefit from the identification of pathological outliers in this work. These cases are highly concentrated in the 5\% of subjects flagged by the algorithmic quality controls, which also helped to identify about 40 suspected cases of renal fusion. 
With median total kidney volumes of \mbox{277 cm$^3$} for men and \mbox{220 cm$^3$} for women, the volumes acquired by the proposed pipeline are smaller than those typically reported in the medical literature, especially when more than just parenchymal tissue is selected.
In comparison, a previous study of 150 men and women reported volumes that were about 35\% larger, based on a disc-summation method in MRI that excluded the renal pelvis and vasculature, with further validation by a water displacement method.\cite{cheong2007normal}
Another study of 1,852 men and women yielded volumes that were about 20\% larger, based on a disc-summation method on manual delineations in MRI that excluded cysts and large vessels.\cite{roseman2017clinical} Values similar to those obtained in this work occur only in their reported lower quartile range of measurements.
More similar values were obtained by previous studies that also focussed on the renal parenchyma, segmenting cortex and medulla only. For segmentations in CT of 1,344 men and women, the renal parenchyma was reported to be about 8\% larger in a subgroup with similar mean age to this work.\cite{wang2014age}
In yet another study with MRI of 50 men and women with renovascular disease, the reported volumes were only about 3\% larger, based on manual segmentations and voxel count measurements in MRI, with only cortical and medullary tissue being included \cite{gandy2007clinical}.
%

In terms of methodology, kidney segmentations with Dice scores of up to 0.974 have been reported in the literature for benchmark challenges involving neural networks on CT data.\cite{isensee2019attempt, heller2019state} Reaching comparable quality on the UK Biobank neck-to-knee body MRI may not be technically feasible, as the given images are of lower resolution and even repeat segmentation by human operators yielded lower consistency in this work. With no fixed image contrast, such as the Hounsfield units in CT, an objectively consistent placement of the kidney outline in MRI is more challenging. Nonetheless, it is possible that a 3D network architecture could reach superior performance. Future work may explore this potential, but will have to account for the massively increased runtime requirements for 3D architectures. Based on the reported runtimes\cite{isensee2019attempt}, a 3D network may require up to an entire day for training as opposed to the 30 minutes for the 2.5 architecture used in this work, and a similar factor may apply to the inference.
Competitive results have also been reported for other approaches that do not utilize neural networks. A recently published approach with appearance-guided deformable boundaries reached a mean Dice score of 0.95 with a 9.5\% percentage error for total kidney volume in abdominal diffusion MRI of 72 men and women\cite{shehata20183d}. Whereas these metrics are similar to the validation results reached in this work, it is worth noting that their reported runtime would also amount to a total of almost two months for inference on 40,000 subjects as compared to the two days required by the proposed segmentation pipeline. In an older technique based on adaptive region growing in CT of 30 subjects, comparable quality was only reached in the best case, with a mean Dice score of 0.88\cite{lin2006computer}. A more recent work with a multi-atlas technique reported a mean Dice of 0.952 for kidney segmentation in CT of 22 subjects\cite{yang2014automatic}. The latter may not be directly comparable to the proposed pipeline however, as a rather convex segmentation style and about double the image resolution available in the UK Biobank was used. Another previous study on CT of ADPKD, segmented with fully convolutional neural networks similar to the one proposed in this work, reported a mean Dice score of 0.86 for three data of three different studies.\cite{sharma2017automatic}

When training the network, dataset \textit{B} was provided for additional guidance on the most challenging morphology. Nonetheless, severe anomalies such as renal fusion and suspected ADPKD were excluded and are thereby effectively accepted as failure cases of the proposed pipeline. This design decision was motivated by the concern that the network may learn a compromise, allowing for better results on these outliers while simultaneously performing worse on the majority of typical cases. The benefit of dataset \textit{B} is not immediately clear from the validation results of Table \ref{tab_validation}, where the main result is actually outperformed by the cross-validation using only single-operator data. This is likely a side-effect of validating on references created by the operator of dataset \textit{A} only. The individual segmentation styles of the two operators of dataset \textit{B} show a tendency towards oversegmentation relative to the operator of dataset \textit{A}, marking about 10\% more volume. This tendency is emulated by the network trained on the combined datasets, which learned a compromise of segmentation styles. This compromise achieves a slightly lower agreement with the references of dataset \textit{A}, but nonetheless appears more robust. Preliminary inference runs, which were trained on dataset \textit{A} only, produced up to three times more \textit{scrap volume} than the main configuration presented here. 

The 2.5D U-Net was able to correctly segment one subject with a missing right kidney in dataset \textit{A} during cross-validation, even though no comparable case existed in the training data. It is possible that the two-dimensional input format may have enabled the network to learn unilateral segmentation based on individual slices containing tissue of only one kidney, with the other being further above or below. Although no rigorous comparison was attempted here, the 2.5D modifications to the original U-Net \cite{ronneberger2015u} are estimated to accelerate training by about 25\% and increase the Dice score by about 0.02.

Several limitations apply. The neural networks trained in this work can only be expected to show comparable performance on future MRI with the same imaging protocol, type of MRI device, and subject demographics. When applied to data of other studies, new training data may be required. The given MRI data is arguably not optimal for kidney volumetry, being originally intended for body composition analysis.\cite{west_feasibility_2016} With kidney tissue being typically contained in two breath-hold imaging stations, the measurement error is potentially compounded by artefacts such as motion and other factors that are not represented in the validation metrics. Even though the algorithmic quality ratings can be expected to identify the worst cases, actual correction may be possible with registration techniques for image mosaicing\cite{ceranka2018registration}, which were not attempted here. Similarly, those cases excluded by the location rating could be trivially recovered by including the adjacent imaging stations in the pipeline. Another limitation is the degree to which the algorithmic quality ratings themselves are automated. With intuitive, rule-based scores, they provide a high level of control over the exclusions and successfully identify the most severe outliers and failure cases. While the need for manual controls is thereby vastly reduced, the study of their distributions and choice of percentiles does require human intervention and would ideally be automated entirely. No guarantee is provided that all failure cases are exhaustively identified, or that the excluded cases are indeed inadequate. The conservative criteria chosen in this work exclude 5\% of cases, which nonetheless translates to about 2,000 subjects for further inspection, many of which are presumably of acceptable quality. The current post-processing steps may furthermore yield misleading results in rare cases where severe cystic formations fragment the healthy kidney tissue such that objective delineation is no longer possible. In these cases, the two largest connected components may occur on the same side, leading to implausible distance and unilateral volume measurements. 

The inference of high-quality measurements will therefore remain a continuous effort. A considerable advantage is posed by the high speed of the proposed pipeline, simplifying future coverage of newly released scans and repeated inference runs. With the latter, differently trained segmentation networks could potentially be applied for inference, with the model variation serving as a proxy for prediction uncertainty. Similarly, reference segmentations that segregate cortex and medulla, include or even isolate cysts may enable the inference of new, dedicated measurements to complement those obtained in this work. With new post-processing steps it would also be possible to provide measurements of kidney length, width, and depth so that ellipsoid volumes could be studied. More elaborate quality controls could furthermore rely on independent shape models or atlas segmentations, which have been previously used for large-scale quality controls of UK Biobank cardiovascular MRI segmentation in a variation of the concept of reverse classification accuracy.\cite{robinson2019automated} 

While the collection of metadata and MRI acquisition by the UK Biobank are still in progress, the obtained measurements can already be provided as return data and used for further research. Whereas the currently available blood biochemistry and urine assays predate the MRI by several years, various fields on body size and composition are already available for association studies. The latter are often based on semi-automated processing of the same neck-to-knee body MRI as used in this work and do not yet cover all released subjects. Recent work on image-based regression with neural networks for biometry\cite{Langner2020} can nonetheless provide accurate approximations months or years before full coverage by the reference methods is achieved, so that many associations could already be studied. Likewise, genetic information is readily available and future work may also examine the repeat imaging visit as planned for another 10,000 subjects. The proposed pipeline is expected to successfully process these images without any need for changes, and the resulting measurements could enable further study of longitudinal effects and disease outcomes associated with changes in parenchymal kidney volume.

\section{Conclusion}

The proposed kidney segmentation pipeline generates fast, accurate, and objective delineations with close agreement to human operators. It was applied to all available UK Biobank neck-to-knee body MRI, with only 5\% of results showing signs of potential measurement failure. Similar performance is expected for future UK Biobank releases, with the remaining results already forming a substantial sample of left and right parenchymal kidney volume measurements that can be shared for further medical research.

\section*{Acknowledgements}

This work was supported by a research grant from the Swedish Heart- Lung Foundation and the Swedish Research Council (2016-01040, 2019-04756) and used the UK Biobank resource under application no. 14237.
Thanks to Lisa Jarl and Paul Hockings for their advice.

\section*{Author contributions statement}
T.L. wrote the main manuscript text and conducted the experiments, 
A.Ö. conducted an early precursor of the single-operator cross-validation\cite{ostling2019automated},
L.M. provided the reference segmentations of dataset \textit{A},
A.K. and D.O. provided the reference segmentations of dataset \textit{B},
D.L, A.W., and L.L. implemented the 2.5D network architecture in a related project,
R.S. and J.K supervised the project and data access from the UK Biobank,
H.A. provided medical expertise.
All authors read and approved the final manuscript.
 
 
\section*{Additional information}

\bibliography{references}

\begin{thebibliography}{10}
\expandafter\ifx\csname url\endcsname\relax
  \def\url#1{\texttt{#1}}\fi
\expandafter\ifx\csname urlprefix\endcsname\relax\def\urlprefix{URL }\fi
\providecommand{\bibinfo}[2]{#2}
\providecommand{\eprint}[2][]{\url{#2}}

\bibitem{sudlow_uk_2015}
\bibinfo{author}{Sudlow, C.} \emph{et~al.}
\newblock \bibinfo{title}{{UK} {Biobank}: {An} {Open} {Access} {Resource} for
  {Identifying} the {Causes} of a {Wide} {Range} of {Complex} {Diseases} of
  {Middle} and {Old} {Age}}.
\newblock \emph{\bibinfo{journal}{PLOS Medicine}}
  \textbf{\bibinfo{volume}{12}}, \bibinfo{pages}{e1001779}
  (\bibinfo{year}{2015}).
\newblock \urlprefix\url{https://dx.plos.org/10.1371/journal.pmed.1001779}.

\bibitem{west_feasibility_2016}
\bibinfo{author}{West, J.} \emph{et~al.}
\newblock \bibinfo{title}{Feasibility of {MR}-{Based} {Body} {Composition}
  {Analysis} in {Large} {Scale} {Population} {Studies}}.
\newblock \emph{\bibinfo{journal}{PLoS ONE}} \textbf{\bibinfo{volume}{11}}
  (\bibinfo{year}{2016}).
\newblock
  \urlprefix\url{https://www.ncbi.nlm.nih.gov/pmc/articles/PMC5035023/}.

\bibitem{emamian1993kidney}
\bibinfo{author}{Emamian, S.~A.}, \bibinfo{author}{Nielsen, M.~B.},
  \bibinfo{author}{Pedersen, J.~F.} \& \bibinfo{author}{Ytte, L.}
\newblock \bibinfo{title}{Kidney dimensions at sonography: correlation with
  age, sex, and habitus in 665 adult volunteers.}
\newblock \emph{\bibinfo{journal}{AJR. American journal of roentgenology}}
  \textbf{\bibinfo{volume}{160}}, \bibinfo{pages}{83--86}
  (\bibinfo{year}{1993}).

\bibitem{glodny2009kidney}
\bibinfo{author}{Glodny, B.} \emph{et~al.}
\newblock \bibinfo{title}{Kidney fusion anomalies revisited: clinical and
  radiological analysis of 209 cases of crossed fused ectopia and horseshoe
  kidney}.
\newblock \emph{\bibinfo{journal}{BJU international}}
  \textbf{\bibinfo{volume}{103}}, \bibinfo{pages}{224--235}
  (\bibinfo{year}{2009}).

\bibitem{sharma2017kidney}
\bibinfo{author}{Sharma, K.} \emph{et~al.}
\newblock \bibinfo{title}{Kidney volume measurement methods for clinical
  studies on autosomal dominant polycystic kidney disease}.
\newblock \emph{\bibinfo{journal}{PloS one}} \textbf{\bibinfo{volume}{12}}
  (\bibinfo{year}{2017}).

\bibitem{hoy2008nephron}
\bibinfo{author}{Hoy, W.~E.} \emph{et~al.}
\newblock \bibinfo{title}{Nephron number, glomerular volume, renal disease and
  hypertension}.
\newblock \emph{\bibinfo{journal}{Current opinion in nephrology and
  hypertension}} \textbf{\bibinfo{volume}{17}}, \bibinfo{pages}{258--265}
  (\bibinfo{year}{2008}).

\bibitem{rossing2002risk}
\bibinfo{author}{Rossing, P.}, \bibinfo{author}{Hougaard, P.} \&
  \bibinfo{author}{Parving, H.-H.}
\newblock \bibinfo{title}{Risk factors for development of incipient and overt
  diabetic nephropathy in type 1 diabetic patients: a 10-year prospective
  observational study}.
\newblock \emph{\bibinfo{journal}{Diabetes care}}
  \textbf{\bibinfo{volume}{25}}, \bibinfo{pages}{859--864}
  (\bibinfo{year}{2002}).

\bibitem{bakker1998vitro}
\bibinfo{author}{Bakker, J.}, \bibinfo{author}{Olree, M.},
  \bibinfo{author}{Kaatee, R.}, \bibinfo{author}{de~Lange, E.~E.} \&
  \bibinfo{author}{Beek, F.~J.}
\newblock \bibinfo{title}{In vitro measurement of kidney size: comparison of
  ultrasonography and mri}.
\newblock \emph{\bibinfo{journal}{Ultrasound in medicine \& biology}}
  \textbf{\bibinfo{volume}{24}}, \bibinfo{pages}{683--688}
  (\bibinfo{year}{1998}).

\bibitem{cheong2007normal}
\bibinfo{author}{Cheong, B.}, \bibinfo{author}{Muthupillai, R.},
  \bibinfo{author}{Rubin, M.~F.} \& \bibinfo{author}{Flamm, S.~D.}
\newblock \bibinfo{title}{Normal values for renal length and volume as measured
  by magnetic resonance imaging}.
\newblock \emph{\bibinfo{journal}{Clinical journal of the American Society of
  Nephrology}} \textbf{\bibinfo{volume}{2}}, \bibinfo{pages}{38--45}
  (\bibinfo{year}{2007}).

\bibitem{lin2006computer}
\bibinfo{author}{Lin, D.-T.}, \bibinfo{author}{Lei, C.-C.} \&
  \bibinfo{author}{Hung, S.-W.}
\newblock \bibinfo{title}{Computer-aided kidney segmentation on abdominal ct
  images}.
\newblock \emph{\bibinfo{journal}{IEEE transactions on information technology
  in biomedicine}} \textbf{\bibinfo{volume}{10}}, \bibinfo{pages}{59--65}
  (\bibinfo{year}{2006}).

\bibitem{ali2007graph}
\bibinfo{author}{Ali, A.~M.}, \bibinfo{author}{Farag, A.~A.} \&
  \bibinfo{author}{El-Baz, A.~S.}
\newblock \bibinfo{title}{Graph cuts framework for kidney segmentation with
  prior shape constraints}.
\newblock In \emph{\bibinfo{booktitle}{International conference on medical
  image computing and computer-assisted intervention}},
  \bibinfo{pages}{384--392} (\bibinfo{organization}{Springer},
  \bibinfo{year}{2007}).

\bibitem{shehata20183d}
\bibinfo{author}{Shehata, M.} \emph{et~al.}
\newblock \bibinfo{title}{3d kidney segmentation from abdominal diffusion mri
  using an appearance-guided deformable boundary}.
\newblock \emph{\bibinfo{journal}{PloS one}} \textbf{\bibinfo{volume}{13}}
  (\bibinfo{year}{2018}).

\bibitem{heller2019state}
\bibinfo{author}{Heller, N.} \emph{et~al.}
\newblock \bibinfo{title}{The state of the art in kidney and kidney tumor
  segmentation in contrast-enhanced ct imaging: Results of the kits19
  challenge}.
\newblock \emph{\bibinfo{journal}{arXiv preprint arXiv:1912.01054}}
  (\bibinfo{year}{2019}).

\bibitem{sharma2017automatic}
\bibinfo{author}{Sharma, K.} \emph{et~al.}
\newblock \bibinfo{title}{Automatic segmentation of kidneys using deep learning
  for total kidney volume quantification in autosomal dominant polycystic
  kidney disease}.
\newblock \emph{\bibinfo{journal}{Scientific reports}}
  \textbf{\bibinfo{volume}{7}}, \bibinfo{pages}{1--10} (\bibinfo{year}{2017}).

\bibitem{long2015fully}
\bibinfo{author}{Long, J.}, \bibinfo{author}{Shelhamer, E.} \&
  \bibinfo{author}{Darrell, T.}
\newblock \bibinfo{title}{Fully convolutional networks for semantic
  segmentation}.
\newblock In \emph{\bibinfo{booktitle}{Proceedings of the IEEE conference on
  computer vision and pattern recognition}}, \bibinfo{pages}{3431--3440}
  (\bibinfo{year}{2015}).

\bibitem{ronneberger2015u}
\bibinfo{author}{Ronneberger, O.}, \bibinfo{author}{Fischer, P.} \&
  \bibinfo{author}{Brox, T.}
\newblock \bibinfo{title}{U-net: Convolutional networks for biomedical image
  segmentation}.
\newblock In \emph{\bibinfo{booktitle}{International Conference on Medical
  image computing and computer-assisted intervention}},
  \bibinfo{pages}{234--241} (\bibinfo{organization}{Springer},
  \bibinfo{year}{2015}).

\bibitem{han2017automatic}
\bibinfo{author}{Han, X.}
\newblock \bibinfo{title}{Automatic liver lesion segmentation using a deep
  convolutional neural network method}.
\newblock \emph{\bibinfo{journal}{arXiv preprint arXiv:1704.07239}}
  (\bibinfo{year}{2017}).

\bibitem{isensee2019attempt}
\bibinfo{author}{Isensee, F.} \& \bibinfo{author}{Maier-Hein, K.~H.}
\newblock \bibinfo{title}{An attempt at beating the 3d u-net}.
\newblock \emph{\bibinfo{journal}{arXiv preprint arXiv:1908.02182}}
  (\bibinfo{year}{2019}).

\bibitem{zheng20183}
\bibinfo{author}{Zheng, Q.}, \bibinfo{author}{Delingette, H.},
  \bibinfo{author}{Duchateau, N.} \& \bibinfo{author}{Ayache, N.}
\newblock \bibinfo{title}{3-d consistent and robust segmentation of cardiac
  images by deep learning with spatial propagation}.
\newblock \emph{\bibinfo{journal}{IEEE transactions on medical imaging}}
  \textbf{\bibinfo{volume}{37}}, \bibinfo{pages}{2137--2148}
  (\bibinfo{year}{2018}).

\bibitem{bai2018automated}
\bibinfo{author}{Bai, W.} \emph{et~al.}
\newblock \bibinfo{title}{Automated cardiovascular magnetic resonance image
  analysis with fully convolutional networks}.
\newblock \emph{\bibinfo{journal}{Journal of Cardiovascular Magnetic
  Resonance}} \textbf{\bibinfo{volume}{20}}, \bibinfo{pages}{65}
  (\bibinfo{year}{2018}).

\bibitem{attar2019quantitative}
\bibinfo{author}{Attar, R.} \emph{et~al.}
\newblock \bibinfo{title}{Quantitative cmr population imaging on 20,000
  subjects of the uk biobank imaging study: Lv/rv quantification pipeline and
  its evaluation}.
\newblock \emph{\bibinfo{journal}{Medical image analysis}}
  \textbf{\bibinfo{volume}{56}}, \bibinfo{pages}{26--42}
  (\bibinfo{year}{2019}).

\bibitem{Langner2020}
\bibinfo{author}{Langner, T.}, \bibinfo{author}{Ahlström, H.} \&
  \bibinfo{author}{Kullberg, J.}
\newblock \bibinfo{title}{Large-scale biometry with interpretable neural
  network regression on uk biobank body mri}.
\newblock \emph{\bibinfo{journal}{arXiv preprint arXiv:2002.06862}}
  (\bibinfo{year}{2020}).

\bibitem{ostling2019automated}
\bibinfo{author}{{\"O}stling, A.}
\newblock \bibinfo{title}{Automated kidney segmentation in magnetic resonance
  imaging using u-net} (\bibinfo{year}{2019}).

\bibitem{malmberg2017smartpaint}
\bibinfo{author}{Malmberg, F.}, \bibinfo{author}{Nordenskj{\"o}ld, R.},
  \bibinfo{author}{Strand, R.} \& \bibinfo{author}{Kullberg, J.}
\newblock \bibinfo{title}{Smartpaint: a tool for interactive segmentation of
  medical volume images}.
\newblock \emph{\bibinfo{journal}{Computer Methods in Biomechanics and
  Biomedical Engineering: Imaging \& Visualization}}
  \textbf{\bibinfo{volume}{5}}, \bibinfo{pages}{36--44} (\bibinfo{year}{2017}).

\bibitem{ternaus}
\bibinfo{author}{Iglovikov, V.} \& \bibinfo{author}{Shvets, A.}
\newblock \bibinfo{title}{Ternausnet: U-net with vgg11 encoder pre-trained on
  imagenet for image segmentation}.
\newblock \emph{\bibinfo{journal}{ArXiv e-prints}}  (\bibinfo{year}{2018}).
\newblock \eprint{1801.05746}.

\bibitem{bilic2019liver}
\bibinfo{author}{Bilic, P.} \emph{et~al.}
\newblock \bibinfo{title}{The liver tumor segmentation benchmark (lits)}.
\newblock \emph{\bibinfo{journal}{arXiv preprint arXiv:1901.04056}}
  (\bibinfo{year}{2019}).

\bibitem{roseman2017clinical}
\bibinfo{author}{Roseman, D.~A.} \emph{et~al.}
\newblock \bibinfo{title}{Clinical associations of total kidney volume: the
  framingham heart study}.
\newblock \emph{\bibinfo{journal}{Nephrology Dialysis Transplantation}}
  \textbf{\bibinfo{volume}{32}}, \bibinfo{pages}{1344--1350}
  (\bibinfo{year}{2017}).

\bibitem{wang2014age}
\bibinfo{author}{Wang, X.} \emph{et~al.}
\newblock \bibinfo{title}{Age, kidney function, and risk factors associate
  differently with cortical and medullary volumes of the kidney}.
\newblock \emph{\bibinfo{journal}{Kidney international}}
  \textbf{\bibinfo{volume}{85}}, \bibinfo{pages}{677--685}
  (\bibinfo{year}{2014}).

\bibitem{gandy2007clinical}
\bibinfo{author}{Gandy, S.}, \bibinfo{author}{Armoogum, K.},
  \bibinfo{author}{Nicholas, R.}, \bibinfo{author}{McLeay, T.} \&
  \bibinfo{author}{Houston, J.}
\newblock \bibinfo{title}{A clinical mri investigation of the relationship
  between kidney volume measurements and renal function in patients with
  renovascular disease}.
\newblock \emph{\bibinfo{journal}{The British journal of radiology}}
  \textbf{\bibinfo{volume}{80}}, \bibinfo{pages}{12--20}
  (\bibinfo{year}{2007}).

\bibitem{yang2014automatic}
\bibinfo{author}{Yang, G.} \emph{et~al.}
\newblock \bibinfo{title}{Automatic kidney segmentation in ct images based on
  multi-atlas image registration}.
\newblock In \emph{\bibinfo{booktitle}{2014 36th Annual International
  Conference of the IEEE Engineering in Medicine and Biology Society}},
  \bibinfo{pages}{5538--5541} (\bibinfo{organization}{IEEE},
  \bibinfo{year}{2014}).

\bibitem{ceranka2018registration}
\bibinfo{author}{Ceranka, J.} \emph{et~al.}
\newblock \bibinfo{title}{Registration strategies for multi-modal whole-body
  mri mosaicing}.
\newblock \emph{\bibinfo{journal}{Magnetic resonance in medicine}}
  \textbf{\bibinfo{volume}{79}}, \bibinfo{pages}{1684--1695}
  (\bibinfo{year}{2018}).

\bibitem{robinson2019automated}
\bibinfo{author}{Robinson, R.} \emph{et~al.}
\newblock \bibinfo{title}{Automated quality control in image segmentation:
  application to the uk biobank cardiovascular magnetic resonance imaging
  study}.
\newblock \emph{\bibinfo{journal}{Journal of Cardiovascular Magnetic
  Resonance}} \textbf{\bibinfo{volume}{21}}, \bibinfo{pages}{18}
  (\bibinfo{year}{2019}).

\end{thebibliography}

\pagebreak

\section{Supplementary Material}

The following material provides additional detail on the experiments and results, with code samples and further documentation on GitHub:
\url{https://github.com/tarolangner/ukb_segmentation}

\subsection{Validation details}

Jaccard indices for the validation experiments are given in Supplementary Table \ref{tab_supp_jaccard}. More detail, including separate evaluations of left and right kidney volumes, as well as distance, are given in Supplementary Table \ref{tab_supp_cv} for the network and Supplementary Table \ref{tab_supp_operators} for human operator variability.

\begin{table*}[h]
	\begin{center}
		\caption{Jaccard indices for network validation and operator variability}
		\label{tab_supp_jaccard}		
		\begin{tabular}{l|cc|cc}
		
			& \multicolumn{2}{c|}{Network validation} & \multicolumn{2}{c}{Human variability} \\
			& Main result & Single-operator & Intra-operator & Inter-operator \\
			\hline

			Jaccard index & 0.916 & 0.917 & 0.927 & 0.852 \\
			N & 64 & 64 & 5 & 5\\		
			\hline
			\hline
			\multicolumn{5}{l}{Validation on N subjects for the neural network and repeat segmentation by human operators.}\\ 
			\multicolumn{5}{l}{ Whereas the single-operator validation is a classical cross-validation on dataset \textit{A}, } \\
			\multicolumn{5}{l}{the main result was obtained by training on samples of both datasets \textit{A} and \textit{B}.} \\		
		\end{tabular}
	\end{center}
\end{table*}

\begin{table*}[h]
	\begin{center}
		\caption{Detailed results of 8-fold cross-validation on 64 subjects}
		\label{tab_supp_cv}		
		\begin{tabular}{l|rrrc|rrrc}
			& \multicolumn{4}{c}{Main result} & \multicolumn{4}{c}{Single-operator}\\
			\hline
			Property & MAE & SMAPE & R$^2$ & LoA & MAE & SMAPE & R$^2$ & LoA\\
			\hline		
			Volume, total & 9.6 cm$^3$ & 3.77\% & 0.950 & (-25.6 to 13.9 cm$^3$) & 
			8.7 cm$^3$ & 3.38\% & 0.951 & (-22.1 to 23.2 cm$^3$) \\
			
			Volume, left & 5.4 cm$^3$ & 4.09\% & 0.954 & (-15.2 to 8.9 cm$^3$) & 
			5.0 cm$^3$ & 3.66\% & 0.950 & (-13.8 to 14.6 cm$^3$) \\
			
			Volume, right & 4.5 cm$^3$ & 6.62\% & 0.968 & (-12.2 to 6.9 cm$^3$) & 
			4.1 cm$^3$ & 6.42\% & 0.971 & (-9.9 to 10.8 cm$^3$) \\
			
			Distance* & 0.3 mm & 0.25\% & 1.000 & (-0.8 to 0.9 mm) & 
			0.3 mm & 0.24\% & 1.000 & (-1.0 to 0.8 mm)\\
			
			\hline
			\hline
			\multicolumn{9}{l}{Validation metrics with mean absolute error (MAE), symmetric mean absolute percentage error (SMAPE),}\\
			\multicolumn{9}{l}{ coefficient of determination (R$^2$) and 95\% limits of agreement (LoA). One subject with a missing right kidney }\\
			\multicolumn{9}{l}{was excluded from the distance measurements and causes high values for the SMAPE metric as an outlier. }\\ 
		\end{tabular}
	\end{center}
\end{table*}

\begin{table*}[h]
	\begin{center}
		\caption{Detailed operator variability on 5 subjects}
		\label{tab_supp_operators}		
		\begin{tabular}{l|rcrc|rcrc}
			& \multicolumn{4}{c}{Intra-operator variability} & \multicolumn{4}{c}{Inter-operator variability}\\
			\hline
			Property & MAE & SMAPE & R$^2$ & LoA & MAE & SMAPE & R$^2$ & LoA \\
			\hline	
			Volume, total & 
			5.6 cm$^3$ & 2.56\% & 0.994 & (-4.5 to 13.4 cm$^3$) & 
			27.0 cm$^3$ & 10.23\% & 0.839 & (-59.0 to 5.0 cm$^3$) \\			
			
			Volume, left & 
			2.0 cm$^3$ & 1.93\% & 0.996 & (-2.1 to 5.4 cm$^3$) & 
			13.0 cm$^3$ & 10.28\% & 0.870 & (-25.1 to -0.92 cm$^3$)\\
			
			Volume, right & 
			3.6 cm$^3$ & 3.15\% & 0.990 & (-2.8 to 8.3 cm$^3$)& 
			14.0 cm$^3$ & 10.18\% & 0.798 & (-34.4 to 6.3 cm$^3$)\\
			
			Distance & 
			0.2 mm & 0.14\% & 1.000 & (-0.6 to 0.4 mm) & 
			0.4 mm & 0.29\% & 1.000 & (0.1 to 0.7 mm)\\
			\hline
			\hline
			\multicolumn{9}{l}{Blinded repeat segmentation by the operator of dataset \textit{A} resulted in the listed intra-operator variability.} \\
			\multicolumn{9}{l}{The inter-operator variability was determined by averaging the measurements per subject for the two} \\
			\multicolumn{9}{l}{ operators of dataset \textit{B} and comparing the result to those of operator \textit{A}.}\\ 
		\end{tabular}
	\end{center}
\end{table*}

\newpage
\subsection{Algorithmic quality ratings}

This section describes the individual algorithmic quality ratings in detail. Supplementary Fig. \ref{fig_exclusion_dist} shows their distribution in the inference run, with example images for failure cases shown in Supplementary Fig. \ref{fig_supp_excluded}.

When the segmentation yields more than just two connected components, the sum of left and right individual volume may be exceeded by the measured total segmented volume. The share of this superfluous \textit{scrap volume} is an important indicator for potential segmentation failure, as spurious segmentation of disjunct areas is a common failure case of many neural network architectures. This effect can also occur in anomalous cases with severe cystic formations that fragment the kidney tissue. In turn, less than two connected components can occur when only one kidney is segmented. This, too, may not necessarily reflect a failure of the network, as prior surgical removal and genetic variations such as fused horseshoe kidneys may arguably lead to only one volume of healthy kidney tissue existing in the image. Anomaly of the imaging process due to undetected water-fat swaps, motion, and non-standard image contrast also affect this behaviour.

In image fusion, the \textit{segmentation fusion cost} is defined as the sum of absolute differences in the overlap of two adjacent stations, based on their voxel-wise binary labels. The \textit{image fusion cost} uses the same term on the voxel-wise water signal intensities and normalizes the sum with the range of existing intensity values to take into account variations in contrast. High values of these terms indicate that the anatomy in both stations does not line up smoothly due to motion or other artefacts.

Another potential problem is misalignment of the subject, so that the kidneys are not fully contained in the second and third imaging station. The \textit{positioning cost} rating quantifies the offset of the kidney centres of mass from the centre line of the image along the longitudinal axis. High values consequently identify those subjects with segmented kidney volume that is placed too high or too low on the bed, potentially reaching beyond the field of view.

The \textit{segmentation smoothness} of the fused binary labels is then rated by summing up the absolute differences between the segmentation volume to a copy of itself, which is shifted by one voxel along the longitudinal axis. Any sudden change in segmentation between adjacent axial slices is thereby penalized, and high values indicate spurious gaps or islands created by the network.

\begin{figure}[H]
	\centering
	\includegraphics[width=0.45\textwidth]{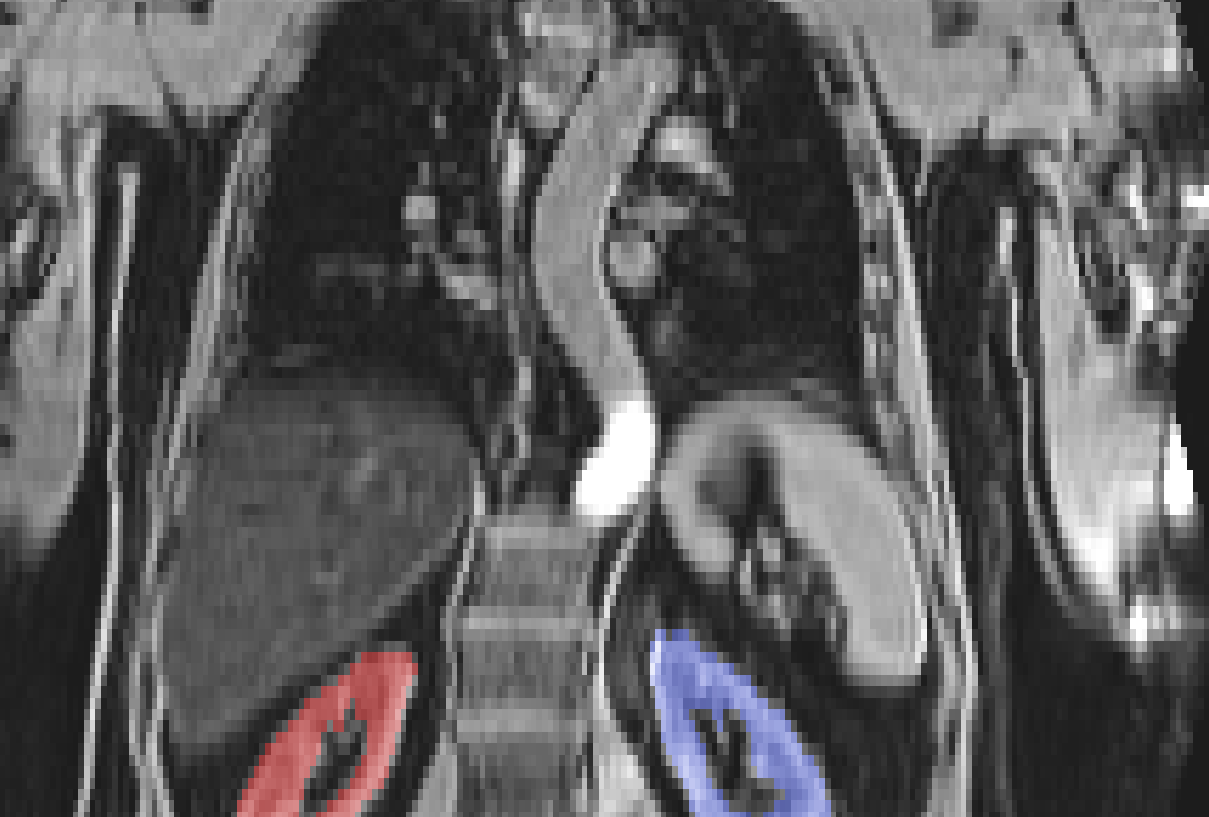}	
	\includegraphics[width=0.45\textwidth]{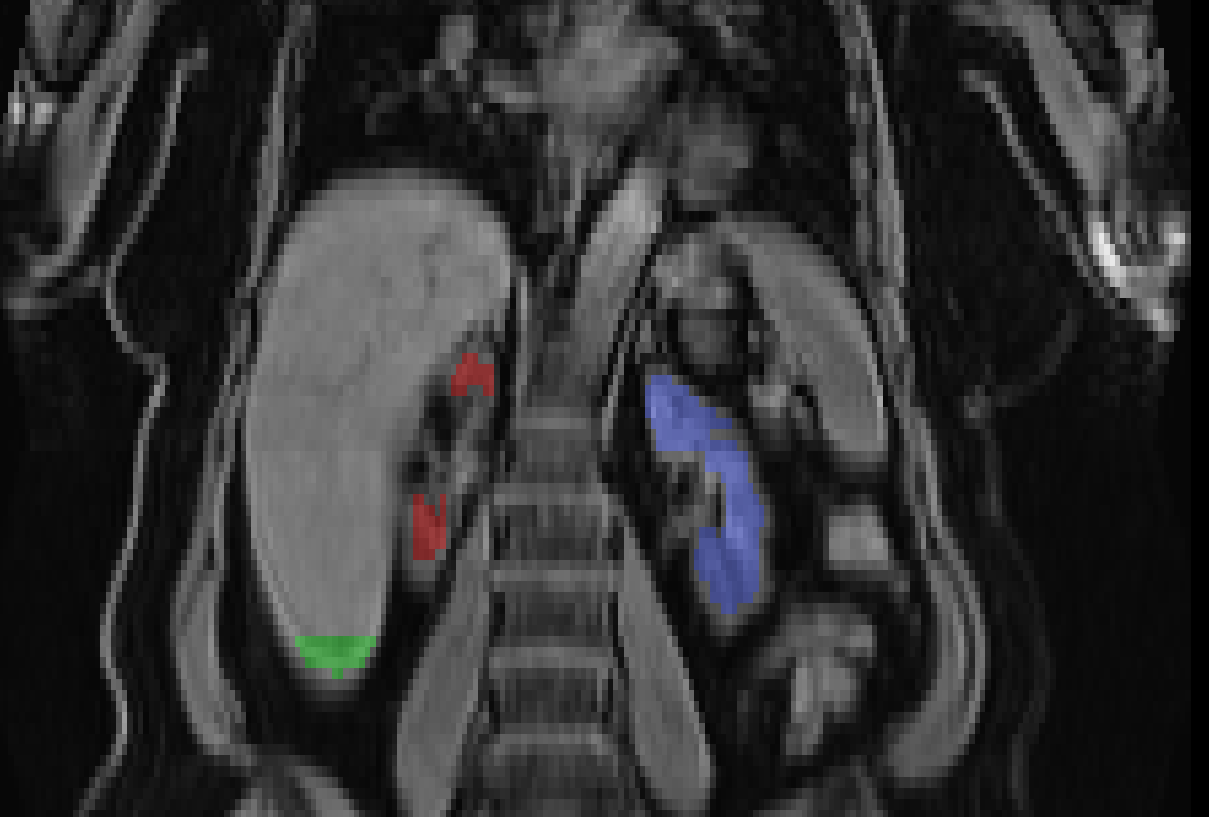}	
	\includegraphics[width=0.45\textwidth]{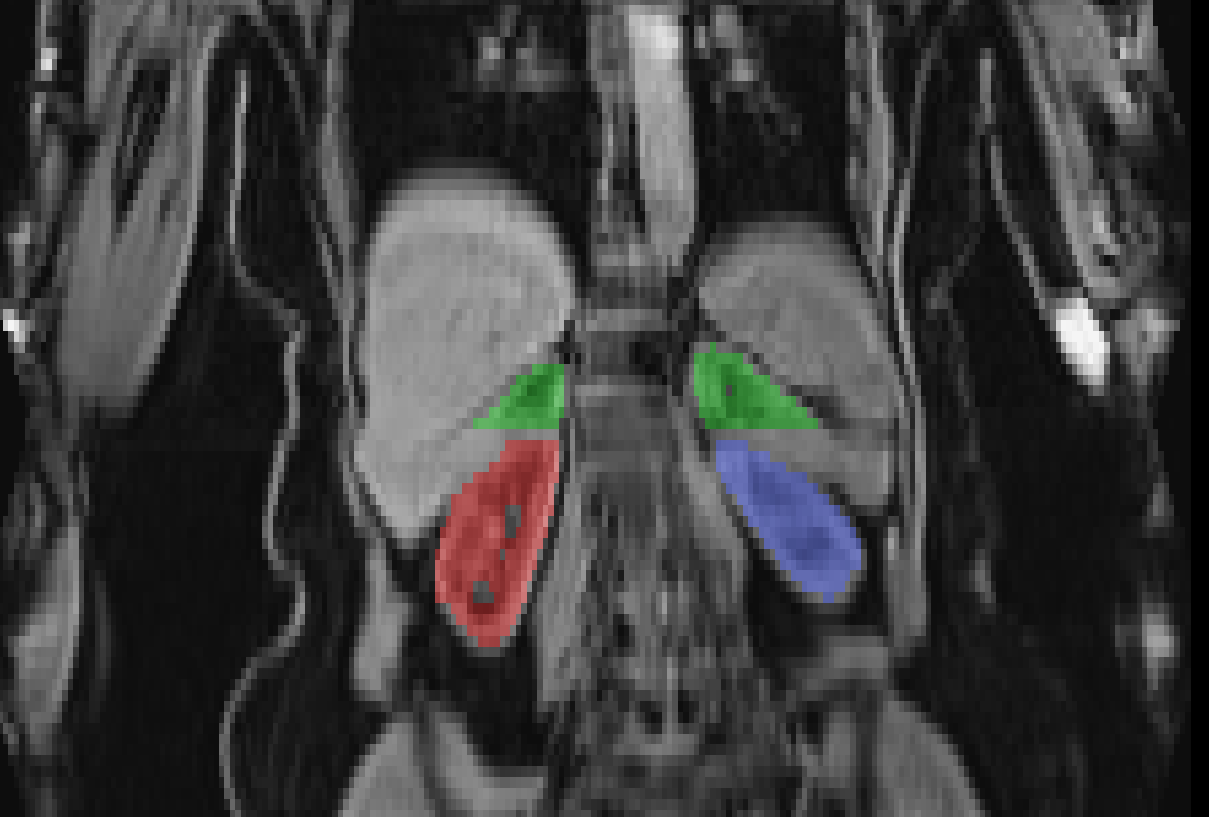}	
	\includegraphics[width=0.45\textwidth]{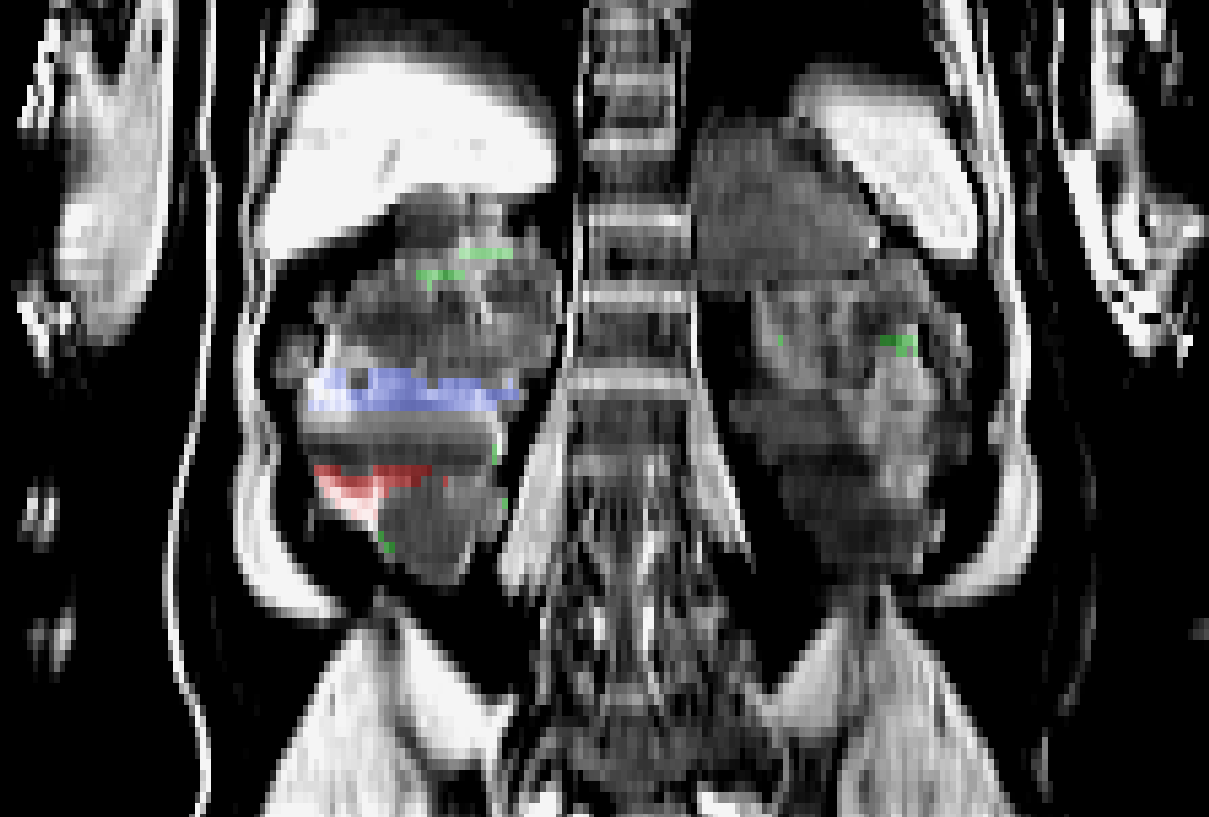}		
	
	\caption{Four characteristic failure cases with segmented right (red), left (blue), and disjunct scrap (green) kidney labels overlaid in coronal view for an outlier in vertical location (top left), genuine segmentation failure, where the network marked parts of the liver (top right), failure due to motion during imaging, despite an arguably correct segmentation (bottom left), and severe fragmentation of healthy tissue by cysts (bottom right).}
	

	
	\label{fig_supp_excluded}	
\end{figure}

\subsection{Kidney offsets}

The left and right kidney are identified as the two largest connected components in the segmented volume. Based on the centre of mass, their relative position in the human body can be described, with euclidean distance values given in Supplementary Table \ref{tab_inference_offset}.

\begin{table*}[h]
	\begin{center}
\caption{Inferred UK Biobank parenchymal kidney distance in mm}
\label{tab_inference_offset}		
\begin{tabular}{l|cl@{\hskip 0.1cm}c|rr@{\hskip 0.2cm}l}
	
	Property & mean $\pm$ SD & [min, & max] & median & (10\%, & 90\%) \\
	\hline	
	\textbf{Distance} &&&&&&\\
	male & 153 $\pm$ 15 & [0, & 211] & 153 & (137, & 171) \\		
	female & 127 $\pm$ 12 & [0, & 215]  & 127 & (114, & 142) \\
	\hline
	\hline
	\multicolumn{7}{l}{Male (N=17,846) and female (N=19,622) distance between left and right centres of mass } \\
	\multicolumn{7}{l}{of parenchymal kidney volume.}
\end{tabular}
	\end{center}
\end{table*}

\end{document}